\documentclass{jltp-num}

\usepackage{graphicx}
\usepackage{amsmath,amssymb}

\newcommand{\bm}[1]{\mbox{\boldmath$#1$}}  
\newcommand{\oN}{f_{_n}}   

\newcommand{\omu}{\tilde{\mu}}
\newcommand{\HO}{\bm{\hat{\cal H}}{}}

\newcommand{\RE}{ \mbox{\boldmath$\Re e$} }

\title{Frequency Dependent Conductivity of the 
Fibonacci--Chain%
\thanks{To appear in {\it Journal of Low Temperature Physics}, March 2002}}

\author{Dieter Walther and Ralph v. Baltz}

\address{Institut f\"ur Theorie der Kondensierten Materie,\\ 
         Universit\"at Karlsruhe, 76128 Karlsruhe, Germany}

\runninghead{D. Walther and R. v. Baltz}{Conductivity of the Fibonacci Chain}

\begin{document}

\maketitle

\begin{abstract} 
A real--space-renormalization method for the frequency dependent conductivity
of the periodic approximants of the Fibonacci chain is developed. 
This scheme is based on the known
$2\times 2$ transfer matrices  and  additional $5\times 5$ matrices
which allow an efficient numerical evaluation of the Kubo formula.
Numerical results are presented.

\end{abstract}

\section{INTRODUCTION}

Quasicrystals (QC)\cite{book} and their periodic approximants reveal 
unusual transport
properties, e.g., extremely low conductivity for alloys of metallic 
constituents\cite{review}, and it is generally  believed that these 
properties are directly related  to the long range but nonperiodic order.
However, no coherent theoretical framework  has been successfully established. 
For instance,  previous calculations of the density of states
(DOS) revealed a set of peaks and gaps of small width 
($10-100$meV)\cite{spikyDOS} but recent studies provide convincing arguments
that these structures are  numerical artifacts\cite{Zilstra}. 
Most of the previous numerical studies of the conductivity were based on the 
Landauer formula\cite{IounePole} but  only few studies were made in the 
framework of linear response theory\cite{BenAbrahamLahiriRoche}. 

One--dimensional QCs, such as the Fibonacci chain (FC), have been studied
in great detail by many researchers, because exact results and
analytical treatments are possible.
The most striking feature of this model is that quasiperiodicity induces long 
range correlations giving rise to an intermediate state of localisation. 
These so--called `critical 
states' are associated to singular continuous spectra of the Hamiltonian\cite{review}, 
while all states are localized for randomly disordered systems in 1--D. 
Most progress in investigating 1--D models has been made by 
methods based upon the trace--map and real--space decimation techniques. 
To the best of our knowledge nothing comparable has been worked out for the 
current--current correlation function which allows for a renormalization group
approach of the frequency dependent conductivity.

Our paper is organized as follows.
Sec.\ 2 summarizes the physical model, in particular, the relation between the
geometrical and the electronic structure which has been  worked
out in a preceding paper\cite{Walther}.
In Sec.\ 3, a suitable expression of the Kubo conductivity is derived
in terms of the known $2\times 2$ transfer matrices and  additional
$5\times 5$ matrices ${\bm \Delta}$, which allow to set up a real--space 
renormalization scheme and its numerical implementation. 
Sec.\ 4  gives some numerical results and conclusions.

\section{FIBONACCI CHAIN}

\begin{figure}
\begin{displaymath}
       \hspace{3em} \mbox{
\unitlength0.35cm
\begin{picture}(30,4)
\thicklines
    \put(0,0){\line(1,0){2}}
                                   \put(0.2,-1){ {\footnotesize $ t_L $ } }
   \put(2,0){\circle*{0.6}}  \put(2,-0.2){\line(1,0){2}}
       \put(1.5,0.7){ {\footnotesize $ \epsilon_{\beta} $ } }
                             \put(2,0.2){\line(1,0){2}}
                                   \put(2.2,-1){ {\footnotesize $ t_S $ } }
   \put(4,0){\circle*{0.6}}  \put(4,0){\line(1,0){2}}
       \put(3.5,0.7){ {\footnotesize $ \epsilon_{\gamma} $ } }
                                   \put(4.2,-1){ {\footnotesize $ t_L $ } }
   \put(6,0){\circle*{0.6}}  \put(6,0){\line(1,0){2}}
       \put(5.5,0.7){ {\footnotesize $ \epsilon_{\alpha} $ } }
                                   \put(6.2,-1){ {\footnotesize $ t_L $ } }
   \put(8,0){\circle*{0.6}}  \put(8,-0.2){\line(1,0){2.2}}
       \put(7.5,0.7){ {\footnotesize $ \epsilon_{\beta} $ } }
                             \put(8,0.2){\line(1,0){2}}
                                   \put(8.2,-1){ {\footnotesize $ t_S $ } }
   \put(10,0){\circle*{0.6}}  \put(10,0){\line(1,0){2}}
       \put(9.5,0.7){ {\footnotesize $ \epsilon_{\gamma} $ } }
                                   \put(10.2,-1){ {\footnotesize $ t_L $ } }
   \put(12,0){\circle*{0.6}}  \put(12,-0.2){\line(1,0){2}}
      \put(11.5,0.7){ {\footnotesize $ \epsilon_{\beta} $ } }
                              \put(12,0.2){\line(1,0){2}}
                                   \put(12.2,-1){ {\footnotesize $ t_S $ } }
   \put(14,0){\circle*{0.6}}  \put(14,0){\line(1,0){2}}
      \put(13.5,0.7){ {\footnotesize $ \epsilon_{\gamma} $ } }
                                   \put(14.2,-1){ {\footnotesize $ t_L $ } }
   \put(16,0){\circle*{0.6}}  \put(16,0){\line(1,0){2}}
      \put(15.5,0.7){ {\footnotesize $ \epsilon_{\alpha} $ } }
                                   \put(16.2,-1){ {\footnotesize $ t_L $ } }
   \put(18,0){\circle*{0.6}}  \put(18,-0.2){\line(1,0){2}}
      \put(17.5,0.7){ {\footnotesize $ \epsilon_{\beta} $ } }
                              \put(18,0.2){\line(1,0){2}}
                                   \put(18.2,-1){ {\footnotesize $ t_S $ } }
   \put(20,0){\circle*{0.6}}  \put(20,0){\line(1,0){2}}
      \put(19.5,0.7){ {\footnotesize $\epsilon_{ \gamma} $ } }
                                   \put(20.2,-1){ {\footnotesize $ t_L $ } }
   \put(22,0){\circle*{0.6}}  \put(22,0){\line(1,0){2}}
      \put(21.5,0.7){ {\footnotesize $ \epsilon_{\alpha} $ } }
                                   \put(22.2,-1){ {\footnotesize $ t_L $ } }
   \put(24,0){\circle*{0.6}}  \put(24,-0.2){\line(1,0){2}}
      \put(23.5,0.7){ {\footnotesize $ \epsilon_{\beta} $ } }
                              \put(24,0.2){\line(1,0){2}}
                                   \put(24.2,-1){ {\footnotesize $ t_S $ } }
\end{picture}
       }
\end{displaymath}
\caption{ Fifth generation standard Fibonacci lattice 
  $w_5 = L S L L S L S L L $ $S L L S $ 
  ($f_5=13$) corresponding to  
  the FC supercell of a periodic approximant.}  
\label{FChain}
\end{figure}
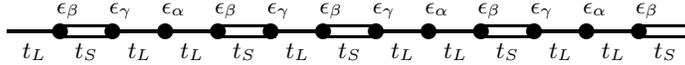

The  FC is specified by two different types of 
``bonds'' denoted by L (``long'') and S (``short''), see Fig.\ 1.
One may describe the geometric structure of such lattices by {\it words} 
$w(L,S)$, i.e. strings in the symbols $L,S$ representing the corresponding 
linear arrangement of the ``letters'' $L$ and $S$. 
These words can be generated by an (infinite) repetition of the recursion law 
(Nielsen transformation (NT))
\begin{equation}
   w_{n} = w_{n-1}  \ast w_{n-2}, \hspace{2em} w_{-1}=S, \quad w_0=L\, ,
   \label{2.2}                           
\end{equation}
where the symbol $\ast$ is defined as the concatenation of two strings and 
$n=1,2,\dots$
The {\it length} of the FC in the $n^{th}$ generation 
$ |w_n| = f_n $, i.e. the sum of symbols $L$ 
and $S$ in $ w_n(L,S) $  satisfies the recursion relation for the 
{\it Fibonacci numbers}\cite{Fib1202}
\begin{equation}
   f_n = f_{n-1} + f_{n-2},  \hspace{2em}  f_{-1}=f_0=1\, .
   \label{2.3}
\end{equation}
Alternatively, the $n^{th}$ order approximants can be generated by substitutions
(morphisms), which operate on the symbols $L,S$ rather than on words  $w_n(L,S)$ 
\begin{equation}
          {\bm {\cal L}}: \; \left\{ \begin{array}{l}
                                L \mapsto LS \\
                                S \mapsto L
                         \end{array}
                         \right. .
          \label{pr5}
\end{equation}
The mapping ${\bm {\cal L}}$ may be viewed as the {\it orbit} of $L$, 
i.e. $w_n = {\bm {\cal L}}^n(L)$:
\begin{displaymath}
          L  \stackrel{{\bm {\cal L}}}{\longmapsto} LS 
             \stackrel{{\bm {\cal L}}}{\longmapsto}
         LSL \stackrel{{\bm {\cal L}}}{\longmapsto} LSLLS 
             \stackrel{{\bm {\cal L}}}{\longmapsto}
          \ldots \stackrel{{\bm {\cal L}}}{\longmapsto} LSLLSLSLLSLLS \cdots
\end{displaymath} 

In our study, we employ the following tight--binding Hamiltonian
\begin{equation}
    {\bm{\cal H}} = \sum_{\mu} | \mu > \epsilon_{\mu} < \mu |
     + \sum_{ \mu,\nu } | \mu > t_{\mu,\nu} < \nu |\, ,
    \label{ham-tb}
\end{equation}
where $ |\mu> $ are Wannier states centered at sites 
$\mu \in {\bm G} = [-M, N-1] \subseteq \mathbb{Z}$,
and the real $\epsilon_{\mu},t_{\mu,\nu}$ denote the site--energies and the 
nearest--neighbour transfer integrals, respectively. ($t_{\mu,\nu}>0)$.
The FC has the speciality that there is no $S$--$S$ bonding, 
hence only three of the possible four combinations occur:
\begin{equation}
\epsilon_{\mu} = \left\{ \begin{array}{r@{\quad:\quad}l}
                             \epsilon_{\alpha} & t_{\mu -1,\mu} =
                                t_{\mu,\mu+1} = t_L \, ,  \\
                             \epsilon_{\beta} & t_{\mu-1,\mu} = t_L, \;
                                t_{\mu,\mu+1} = t_S \,,  \\
                             \epsilon_{\gamma} & t_{\mu-1,\mu} = t_S, \;
                                t_{\mu,\mu+1} = t_L \,.
                             \end{array} \right.  
\label{2.7}  
\end{equation}

To each word $w = (y_{\mu})_{\mu \in G} \equiv y_{_{-M}} y_{_{-M+1}} \cdots y_{_{N}}$,
with $y_{\mu} \in \{ L, S \}$,  $M + N \ge2$, we uniquely assign  a {\it dual word}
$\Sigma = ( \sigma_{\mu} )_{\mu=-M}^{N-1} \equiv (\sigma_{_{-M}},
\ldots , \sigma_{_{N-1}})$, where each symbol 
$\sigma_{\mu} \in \{ \alpha,\beta,\gamma \}$
will be related to the pair $(y_{\mu},y_{\mu +1})$ by the map $ ( L L, L S, S L )
\leftrightarrow ( \alpha,\beta,\gamma )$.
For example,  the dual word pertaining to $w_5$ is 
$ \Sigma^G = \gamma \beta \gamma \alpha \beta \gamma \beta \gamma
\alpha \beta \gamma \alpha$, with ${\bm G} = [0, 12]$, see Fig.\ \ref{FChain}. 
Let be $\Sigma_{u}= (\sigma_{\mu}^u )_{\mu=-m}^{n-1}$ and 
$\Sigma_{v} = (\sigma_{\mu}^v )_{\mu=-p}^{q-1}$ the dual words of $u$, and $v$ 
respectively. We define the product of two dual words, corresponding to
the product $w = u \ast v$, by $\Sigma_{u} \wedge \Sigma_{v} =: \Sigma_{w}$, with 
$\Sigma_{w}= ( \sigma_{- m}^u, \cdots \sigma_{n-1}^u, \sigma_{n}, \sigma_{-p}^v, 
\cdots \sigma_{q-1}^v)$,
$ \; \sigma_n := (y_{n}^u, y_{-p}^v)$, where $y_{n}^u$ ($y_{-p}^v$) is the last (first)
letter of $u$ ($v$).

To set--up a renormalization scheme we approximate the aperiodic FC by  
a periodic approximant of length $2N$ ($M=N$),
\begin{equation}
        \Sigma^{_{^G}} = \underbrace{\Sigma_n \wedge \Sigma_n \wedge \ldots 
                                     \wedge \Sigma_n}_{ 2 N^{\prime} 
        \mbox{\scriptsize times}} \; , 
        \hspace{1em} {\bm G} = [-N, N-1], \;\; N = N^{\prime} \oN\, ,
        \label{pr10_1}
\end{equation} 
where $\oN - 1$ denotes the number of symbols $\sigma_{\mu}$ of $\Sigma_n$.
$\epsilon_{^{_{\! \mu+\oN}}} = \epsilon_{^{_{\! \mu }}}$,
$t_{^{_{\! \mu+\oN}}} = t_{^{_{\! \mu}}}$,
$\mu = \omu + m \oN$ where $\omu \in [0,\oN -1]$
labels the `atoms' inside the Fibonacci supercell
and $m \in [-N^{\prime},N^{\prime}-1]$ labels the vectors of the Bravais lattice
of the periodic approximant. 
Obviously, there is a unique correspondence between the dual word 
$\Sigma$ and the FC Hamiltonian $\HO_{_{\Sigma}}$.

In site representation, the eigenstates of Eq.\ (\ref{ham-tb})
can be represented in terms of the two fundamental solutions 
$\{ {\cal P}_{\mu} \}$  and $\{ P_{\mu} \}$  of 
the discrete Schr\"odinger equation
\begin{equation}
    \big(\HO_{{_{\Sigma_n}}} \psi\big)_{\mu} :=t_{\mu+1}\Psi_{\mu+1}(E) + 
     t_{\mu}\Psi_{\mu-1}(E) + \epsilon_\mu\Psi_{\mu}(E) = E\, \Psi_{\mu}(E)\, ,
\label{Schroed-eq}
\end{equation}
such that the (unnormalized) solutions are
\begin{equation}
      \Psi_{\mu}^{\pm}(E) := {\cal P}_{\mu}(E) +  m_{\pm}(z)P_{\mu}(E) \, .
      \label{pr14}
\end{equation}
$P_{\mu}(E)$, ${\cal P}_{\mu}(E)$ respectively denote polynomials of degree $\mu$ and
$\mu-1$ with initial conditions $P_{_{-1}} = 0$, $P_{_0} = 1/t_{_0}$ 
and ${\cal P}_{_{-1}} = -1/t_{_0}$, ${\cal P}_{_0}= 1$\cite{O-poly}.
In addition, we have to fulfill the Bloch conditions
$\Psi_{\mu \pm \oN} = \exp(\pm\imath k a) \Psi_{\mu}$, where $ a=\oN$ is the ``lattice
constant'' and $k=k(E)$ is the wave vector. According to time--reversal symmetry a
state of energy $E$ is a twofold degenerate which will be denoted by $|E,\pm\rangle$
and $\Psi^\pm_\mu(E)=\langle\mu|\pm\rangle$, respectively.
$m_{\pm}(E)$ denotes the Titchmarsh--Weyl function\cite{TitWeyl} 
\begin{eqnarray}
      m_{\pm}(E)&=& \frac{\xi_{\pm}(E) + t_{_0} {\cal P}_{_{\oN -1}}(E)}
                         {t_{_0} P_{_{\oN -1}}(E)}\,,\\
      \xi_{\pm}(E) &:=& e^{\pm i k(E) \oN} = 
        \frac{\Delta_{_{\! \oN}}}{2} \pm \sqrt{
        \left( \frac{\Delta_{_{\! \oN}}}{2} \right)^2 - 1} \, \, ,\\
       \Delta_{_{\! \oN}} &=& t_{_0} (P_{_{\! \oN}} - {\cal P}_{_{\! \oN -1}})\,,
      \label{pr15}
\end{eqnarray}
where 
$\oN k(E) =(p \pi)/N^{\prime}= \arccos\!\big(\Delta_{_{\oN}}(E)/2\big) \in [0,\pi)$,
and $ 0 \le p < N^{\prime}$.

For the standard one-atomic tight binding chain with equal $t's$ and $\epsilon's$
$\Delta_{_{\! \oN}}\!(E)$ and $P_{\mu}(E) = {\cal P}_{\mu -1}$  are, apart from a scale factor,  Chebyshev polynomials
$T_\mu(x)$, $U_\mu(x)$ of first and second kind, respectively\cite{Abramowitz}.
$\Psi_{\mu}^{\pm}(E) = e^{\pm \imath \mu  k}$, $E(k)=\epsilon -2t\cos(k)$.

\section{CONDUCTIVIY AND REAL--SPACE RENORMALIZATION}

In the linear response theory the real part of the conductivity is given 
by\cite{linres}
\begin{eqnarray}
    \RE\{ \sigma^{(n)}(\omega) \}  &=&   
       \frac{\pi}{\hbar} \int_{- \infty}^{\infty} 
       \frac{f(E)-f(E+\hbar\omega)}{\hbar\omega} \,
       \Gamma^{(n)}(E,\omega ) \, \mbox{d}E \, ,
    \label{pr16}\\
   \Gamma^{(n)}(E,\omega ) &=& \mbox{\bf Tr} \big\{ {\bm{\hat J}} \, 
   \delta \big( E - \HO_{^{_{\Sigma^{(n)}}}} \big) {\bm{\hat J}} \,
   \delta \big( E + \hbar \omega - \HO_{^{_{\Sigma^{(n)}}}}\big) \big\}\, ,
   \label{pr17}
\end{eqnarray}
where $f(E)$ is the Fermi function and $\bm{\hat J}$ is the current operator 
\begin{equation}
       {\bm{\hat J}} = \frac{i e \hbar}{m}\sum_{\mu, \nu \in G}
                       j_{\mu \nu} \big| \mu \big> \big< \nu \big|, \hspace{1em}
       j_{\mu \nu} = - t_{\mu +1} \delta_{\mu \nu-1} + t_{\mu} \delta_{\mu \nu+1}\,.
\end{equation}

\begin{figure}[t]
  \centerline{\includegraphics[width=5in,angle=0]{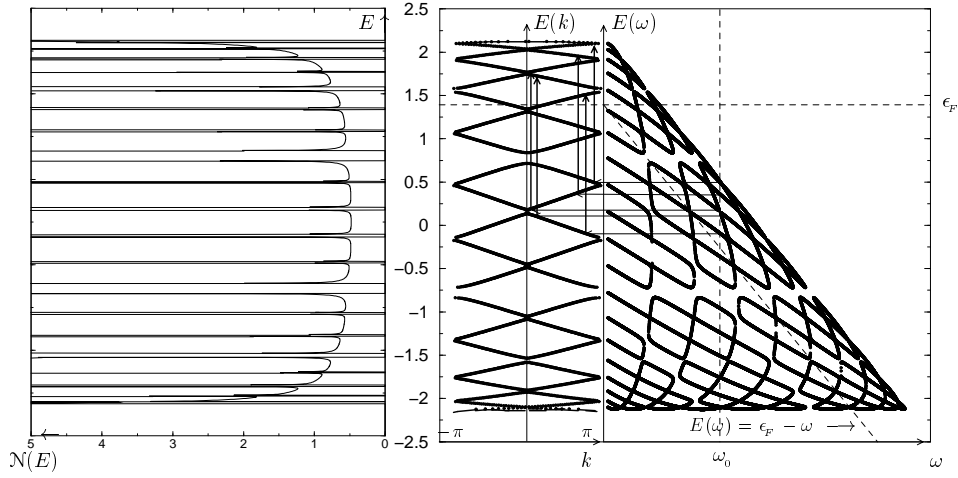}}

  \caption{Density of states (left), bandstructure (middle), 
           and (unrestricted) solutions of Eq.~(\ref{tr-en})
           (right) as function of frequency.
           $6^{th}$ order generation FC with  
           $\epsilon_{\alpha} = \epsilon_{\beta} = \epsilon_{\gamma} = 0$ 
           and $t_{_{\! L}} = 1.1, \: t_{_{\! S}} = 1.0$.}
  \label{DOS-fig} 
\end{figure}

At zero temperature and 
after some nontrivial algebra Eq.\ (\ref{pr16}) becomes
\begin{equation}
      \sigma^{(n)}(\omega) 
                 = \frac{- e^2 }{\hbar^2 \oN} \, 
                     \sum_{E_{\omega}} \left. 
                          \frac{\varsigma \, 
                            \mbox{\bf Tr} \Big\{ 
          \overline{{\bm T}} {\bm \Theta} {\bm T} \overline{{\bm \Theta}} 
  +  \overline{{\bm T}}^{-1} {\bm \Theta} {\bm T}^{-1} \overline{{\bm \Theta}} 
  - 2 {\bm \Theta}{\bm \Theta}
                            \Big\} 
                          }
  { 2 \omega \big| {\overline{\Delta}}{}^{\prime}_{_{\! \oN}} 
                                       - \Delta{}^{\prime}_{_{\! \oN}} \big| \: 
                            \sqrt{ 4 - \big( \Delta_{_{^{\oN}}}(E)\big)^2 } } 
    \right|_{\mbox{\scriptsize $\begin{array}{l} E = E_{\omega}, \\ \omega > 0 
                                \end{array}$}}
\label{pr20}
\end{equation} 
with $\Delta^{\prime} := \mbox{d}\Delta/\mbox{d}E$ and 
$\varsigma=\mbox{sign}(P_{_{\! \oN-1}}(E)P_{_{\! \oN-1}}(E+\hbar\omega))$, 
respectively. (Note, in Eq. (\ref{pr20}) we have omitted the dc--part 
$g^{(n)} \delta(\omega)$ which vanishes in the limit $n\to\infty$). 
All quantities are functions of $E$ and/or $E+\hbar \omega$ and 
$\overline{{\bm \Theta}}(E,E+\hbar\omega)={\bm \Theta}(E+\hbar\omega,E)$, 
$\overline{{\bm T}}(E)={\bm T}(E+\hbar\omega)$, etc\dots . 

$E_\omega$ denotes the energy of an initial occupied (lower) band
from which transitions to an empty (upper) band can take place. 
By conservation of energy and (crystal--) momentum
$E=E_\omega$ is a solution of
\begin{equation}
  k(E + \hbar\omega) = k(E)\, \Longleftrightarrow 
\Delta_{_{\! \oN}}(E+\hbar \omega) - \Delta_{_{\! \oN}}(E) = 0 \, .
\label{tr-en}
\end{equation}
Remarkably, the set of all solutions of Eq.\ (\ref{tr-en}), denoted by 
$E(\omega)$, also includes unphysical solutions
which smoothly join $E_\omega$, when both $E+\hbar\omega$
and $E$ are in the gap regions (where $k$ is imaginary), 
see Fig.\ \ref{DOS-fig}. 
Of course, only the subset $E_\omega$ in the spectrum of $\HO_{_{\Sigma_n}}$ 
should  be included in Eq.~(\ref{pr20}).

\begin{figure}[t]
\centerline{\includegraphics[width=5in]{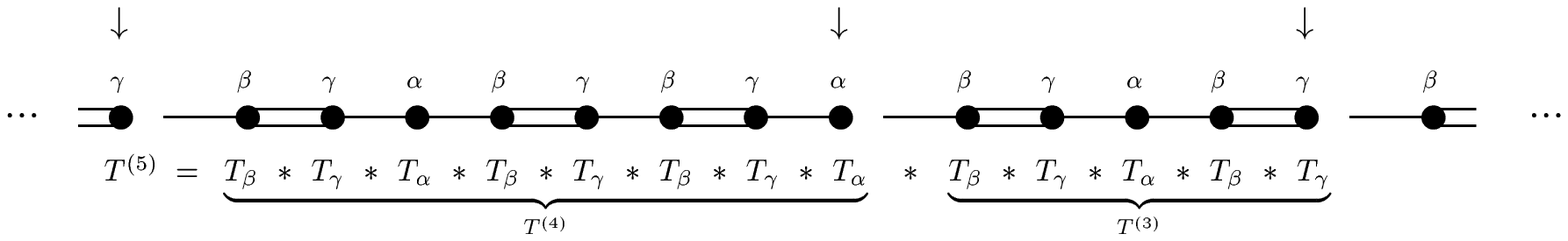}}
\centerline{\includegraphics[width=5in]{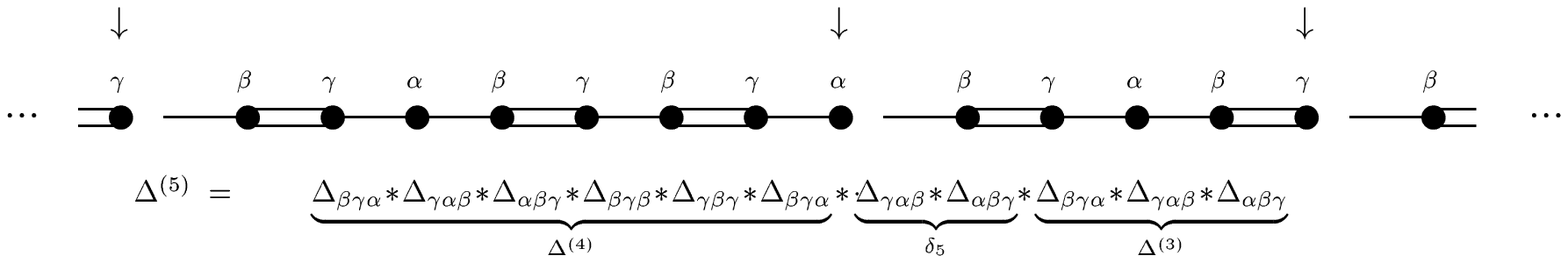}}

\caption{Example of the RTs (\ref{RT_TM}) and (\ref{RT-sandw}) for a 
         $5^{th}$ generation FC.
         Arrows indicate the dual letters $\sigma = \sigma_{_{\! \oN}}^{(n)}$
         as given by Eq.(\ref{Sig-NT}).
         ${\bm T}_1 \ast {\bm T}_2 \ast {\bm T}_3 \equiv 
          {\bm T}_3 \cdot {\bm T}_2 \cdot {\bm T}_1$. etc.,
         where the dot denotes the usual matrix--product.} 

\label{RG-fig}
\end{figure}

To evaluate Eq.\ (\ref{pr20}) we introduce a matrix notation 
\begin{equation}
     {\bm T}(E) \equiv {\bm T}^{(n)}   
                  =    t_{_1} \left( 
      \begin{array}{cc} 
           P_{^{_{\! \oN}}}^{_{^{(1)}}}(E) 
        &  P_{^{_{\! \oN-1}}}^{_{^{(1)}}}(E) \\ 
         - P_{^{_{\! \oN}}}^{_{^{(2)}}}(E)  
        &- P_{^{_{\! \oN-1}}}^{_{^{(2)}}}(E)
      \end{array} \right)\, . 
     \label{pr24}
\end{equation}
$P^{_{^{(\nu)}}}_{^{_{\mu}}}$ are the polynomials associated to $P_{\mu}$
which are obtained by replacing  $\epsilon_{\mu}$, $t_{\mu}$ by
$\epsilon_{\mu+\nu}$, $t_{\mu + \nu}$, e.g. ${\cal P}_{\mu} = P_{^{_{\mu-1}}}^{_{^{(1)}}}$.
In addition,  we define 
${\bm \Theta}(E,E+\hbar \omega) \equiv  {\bm \Theta}^{(n)}$ with
\begin{equation}
     {\bm \Theta}^{(n)} 
             = \frac{1}{t_{_1}} \left(
              \begin{matrix} 
                 2 t_{_1} & E-\epsilon_{_1} \\ 
                 E-\epsilon_{_0} & 2 t_{_1}
              \end{matrix}
              \right)
              \cdot 
              \left(
              \begin{matrix} 
             \Delta^{_{^{\!(n)}}}_{^{_{1 1}}} + \Delta^{_{^{\!(n)}}}_{^{_{1 2}}} 
               & \Delta^{_{^{\!(n)}}}_{^{_{1 4}}} \\ 
                 \Delta^{_{^{\!(n)}}}_{^{_{1 5}}} 
               & \Delta^{_{^{\!(n)}}}_{^{_{1 1}}}
              \end{matrix}
              \right)\, .
    \label{pr25}
\end{equation}
$\Delta^{_{^{\!(n)}}}_{^{_{i j}}}$ are components of the $5 \times 5$--matrix
\begin{equation}
       {\bm \Delta}^{\!\! (n)}(E,E + \hbar \omega)%
                     = {\bm \Delta}_{^{_{\oN-2}}} \, \cdot \,%
                       {\bm \Delta}_{^{_{\oN-3}}} \, \cdot \ldots \cdot%
                       {\bm \Delta}_{^{_{1}}}\, ,%
       \label{pr99}
\end{equation}
where
\begin{equation}
       {\bm \Delta}_{\mu}(E,E + \hbar \omega)  = 
           \left(
              \begin{matrix}
                \overline{s}_{\mu} \xi_{\mu} + 1 & \tau_{\mu} \phi_{\mu} 
               -\overline{s}_{\mu} \xi_{\mu} 
              &-\tau_{\mu} \phi_{\mu} & -\overline{s}_{\mu} \phi_{\mu} 
              & -\tau_{\mu} \xi_{\mu}\\
            1 & 0 & 0 & 0 & 0\\
            0 & 1 & 0 & 0 & 0\\
                \overline{s}_{\mu} & -\overline{s}_{\mu} & 0 & 0 & -\tau_{\mu}\\
                \xi_{\mu} & -\xi_{\mu} & 0 & -\phi_{\mu} & 0  
              \end{matrix}
            \right)\, .
\end{equation}
We have set $s_{\mu} = (E-\epsilon_{\mu})/t_{\mu+1}$, 
$\overline{s}_{\mu} = (E + \hbar \omega - \epsilon_{\mu})/t_{\mu+1}$
$\tau_{\mu} = t_{\mu}/t_{\mu+1}$, and 
$\zeta_{\mu} = s_{\mu} t_{\mu} s_{\mu+1} t_{\mu+2} - 4t_{\mu} t_{\mu+1}$. 
Then $\xi_{\mu} = (\zeta_{\mu} s_{\mu-1} t_{\mu-1} + \zeta_{\mu-1} \tau_{\mu} 
s_{\mu+1} t_{\mu+2} )/$ $(2\zeta_{\mu-1} t_{\mu}) $ and 
$\phi_{\mu} = \zeta_{\mu} \tau_{\mu-1}/\zeta_{\mu-1}$.
In contrast to the matrix 
\begin{equation}
    {\bm T}^{(n)} = {\bm T}_{\! \sigma_{_{\! \oN}}} \cdot 
                      {\bm T}_{\! \sigma_{_{\! \oN-1}}} 
                       \cdot \ldots \cdot  
                      {\bm T}_{\! \sigma_{_{\! 1}}}, \; \mbox{ with }
    {\bm T}_{\! \sigma_{_{\! \mu}}} =  \frac{1}{t_{\mu+1}} 
                \left(
                    \begin{matrix} 
                             0 & 1 \\ 
                      -t_{\mu} & E-\epsilon_{\mu}
                    \end{matrix}
                 \right)\, , 
    \label{pr100}
\end{equation}
which corresponds to each site $\sigma_{\!\mu}$ 
(i.e. to each ``dual letter'', such that 
${\bm T}_{\mu} = {\bm T}_{\sigma_{\mu}}$) 
uniquely, the ${\bm \Delta}_{\mu}$ correlates {\it three} sites 
(i.e. a triple of three dual letters) $\sigma_{\mu-1}, \sigma_{\mu}$ 
and $\sigma_{\mu+1}$, respectively. 
Thus we may write 
${\bm \Delta}_{\mu} ={\bm \Delta}_{\sigma_{\mu-1} 
 \sigma_{\mu} \sigma_{\mu+1} }$, see Fig.\ \ref{RG-fig}. 

The self similarity of FC  shows up in the structure of  
Eqs.\ (\ref{pr99},\ref{pr100}) 
which will lead to a real--space renormalization scheme of the structure 
of the trace--map method\cite{Kohmoto}
or path renormalization used for the calculation of the local DOS\cite{Walther}.
To set up the renormalization scheme we consider the NT of words, 
cf. Eq.\ (\ref{2.2}),
\begin{equation}  
      {\bm X}_{1-s,s} ( w_{k-1},w_{k-2} ) =  ( w_{k},w_{k-1} )\, ,
      \quad
      w_{k} = (w_{k-1})^{1-s} \: w_{k-2} \: (w_{k-1})^{s}\, ,
\label{NT1}
\end{equation}
with $s = 0, 1$, and the initial words $ w_{-1}=S$, $w_0=L $. 
Next, we write the trace--map\cite{Kohmoto} in the form
(recall, that $\Delta_{_{\! \oN}} = \mbox{\bf Tr}\{ {\bm T} \}$)
\begin{eqnarray}
      &{^R}\!{\bm L}(\Delta_{_{\! f_{_{k}} }},\Delta_{_{\! f_{_{k-1}}}},
                     \Delta_{_{\! f_{_{k-1}}}}) 
                       = {^R}\!{\bm R}(\Delta_{_{\! f_{_{k}}}},
                     \Delta_{_{\! f_{_{k-1}}}},\Delta_{_{\! f_{_{k-2}}}}) 
                       = (\Delta_{_{\! f_{_{k+1}}}},\Delta_{_{\! f_{_{k}}}},
                     \Delta_{_{\! f_{_{k-1}}}}),&
      \nonumber \\
      &\Delta_{_{\! f_{_{k+1}}}} = \Delta_{_{\! f_{_{k}}}} 
       \Delta_{_{\! f_{_{k-1}}}} - \Delta_{_{\! f_{_{k-2}}}}\, ,&
      \label{pr32}
\end{eqnarray}
where ${\bm L} = {\bm X}_{1,0}, \; {\bm R} = {\bm X}_{0,1}$. Note, ${^R}\!{\bm L}$ is the renormalization
transformation (RT) corresponding to the NT ${\bm L}$ etc.. Furthermore, the quantity 
$\Lambda =  \Delta_{^{_{f_{_{k-2}}}}}^2 + \Delta_{^{_{f_{_{k-1}}}}}^2 + 
\Delta_{^{_{f_{_{k}}}}}^2 - \Delta_{^{_{f_{_{k-2}}}}}
\Delta_{^{_{f_{_{k-1}}}}} \Delta_{^{_{f_{_{k}}}}}$ 
is invariant under the transformation ${^R}\!{\bm L}$ so that
the trace--map Eq.\ (\ref{pr32}) represents a dynamical system which is 
essentially two--dimensional.
Therefore,
$(\Delta_{^{_{f_{_{n+1}}}}},\Delta_{^{_{f_{_{n}}}}}) = 
\big( {^R}\!{\bm L} \big)^n (\Delta_{^{_{f_{_{2}}}}},\Delta_{^{_{f_{_{1}}}}})$. 

In addition, the NTs (\ref{NT1}) implies
\begin{equation}
     {^R}\!{\bm X}_{1-s_{k},s_{k}}\Big( {\bm T}^{(k-1)}, {\bm T}^{(k-2)}  \Big) 
            = \Big(\Big( {\bm T}^{(k-1)} \Big)^{s_k} {\bm T}^{(k-2)} 
              \Big( {\bm T}^{(k-1)} \Big)^{1-s_k} , {\bm T}^{(k-1)}  \Big) \, . 
    \label{RT_TM}
\end{equation}
By inspection of Eq.\ (\ref{pr100}) we recognize that  ${\bm T}^{(n)}$ 
does not correspond to $\Sigma_{n}$ but to the cyclic permutated word  
${\bm{\cal T}}(\Sigma_{n}) = (\sigma_{_1}, \ldots , \sigma_{_{\oN}}), \;
\sigma_{_{\oN}} = \sigma_{_{0}}$. 
The corresponding word is\cite{Walther1} 
$w^{(n)}(L,S) = {\bm{\cal R}} \circ ({\bm{\cal L}})^{n-1}(L)$ 
which implies
$\big( {\bm T}^{(n+1)},{\bm T}^{(n)}\big) = 
 \big( {^R}\!{\bm L} \big)^{n-1} \circ {^R}\!{\bm R} 
\: \big({\bm T}^{(2)}, {\bm T}^{(1)}\big)$.

The case of the matrices ${\bm \Delta}^{\!(n)}$ is a little bit more 
involved, see Fig.\ \ref{RG-fig}. 
Let ${\bm \Delta}^{\!(n-1)}$ and ${\bm \Delta}^{\!(n-2)}$ 
correspond to the dual words $\Sigma^{(n-1)}$ and $\Sigma^{(n-2)}$,
respectively. 
Then the NT ${\bm L}$ implies
${\bm L}(\Sigma_{n-1},\Sigma_{n-2}) = (\Sigma_{n},\Sigma_{n-1})$, where  
\begin{equation}
      \Sigma_{n} = \Sigma_{n-1} \wedge \Sigma_{n-2} 
                \equiv \left( \sigma_0^{(n-1)}, \cdots , 
                \sigma_{f_{_{n-1}}-1}^{(n-1)}, \sigma_{f_{_{n}}}^{(n)} 
                \sigma_0^{(n-2)}, \cdots , \sigma_{f_{_{n-2}}-1}^{(n-2)} \right)\, .
\label{Sig-NT}
\end{equation}
Eq.\ (\ref{Sig-NT}) corresponds to
$ {\bm \Delta}^{\!(n)} = {\bm \Delta}^{\!(n-2)} \wedge {\bm \Delta}^{(n-1)}
:= {\bm \Delta}^{\!(n-2)} \cdot {\bm \delta}_n \cdot {\bm \Delta}^{\!(n-1)}$, 
where we have introduced a ``sandwich''--matrix
\begin{equation}
      {\bm \delta}_n = {\bm \Delta}_{\sigma_{f_{_{n-2}}-1}^{(n-2)} 
                                     \sigma_0^{(n-1)} \sigma_1^{(n-1)} } \cdot
      {\bm \Delta}_{\sigma_{f_{_{n-2}}}^{(n-2)} 
                    \sigma_{f_{_{n-2}}-1}^{(n-2)} \sigma_0^{(n-1)} }\, .
\end{equation} 
As a result, the NT ${\bm L}$ leads to the following RT for the ${\bm \Delta}^{\!(k)}$
\begin{eqnarray}
      &{^R}\!{\bm L}\Big( {\bm \Delta}^{\!(k-1)},{\bm \Delta}^{\!(k-2)} \Big) =  
                    \Big( {\bm \Delta}^{\!(k-2)} \cdot {\bm \delta}_k \cdot 
                          {\bm \Delta}^{\!(k-1)} , {\bm \Delta}^{\!(k-1)} \Big)\, ,
      & \label{RT-delta} \\ 
      &{\bm \delta}_k = 
      \left\{ 
       \begin{array}{l}  
          {\bm \Delta}_{\gamma \beta \gamma } \cdot 
          {\bm \Delta}_{\beta \gamma \beta  }\, , \;\; k \; \mbox{ even } \\
          {\bm \Delta}_{\alpha \beta \gamma } \cdot 
          {\bm \Delta}_{\gamma \alpha \beta }\, , \;\; k \; \mbox{ odd }
       \end{array} 
      \right. 
      \label{RT-sandw} ,&
\end{eqnarray}
so that 
$( {\bm \Delta}^{\!(n+1)},{\bm \Delta}^{\!(n)} ) = 
 ( {^R}\!{\bm L} )^n ( {\bm \Delta}^{\!(2)},{\bm \Delta}^{\!(1)})$
holds. Eqs.\ (\ref{pr32},\ref{RT_TM},\ref{RT-sandw}) provides a simple renormalization 
scheme to calculate the frequency dependent conductivity of the standard FC.
This is our main result.

\section{RESULTS AND CONCLUSIONS}

\begin{figure}[t!]
  \centerline{\includegraphics[width=5in]{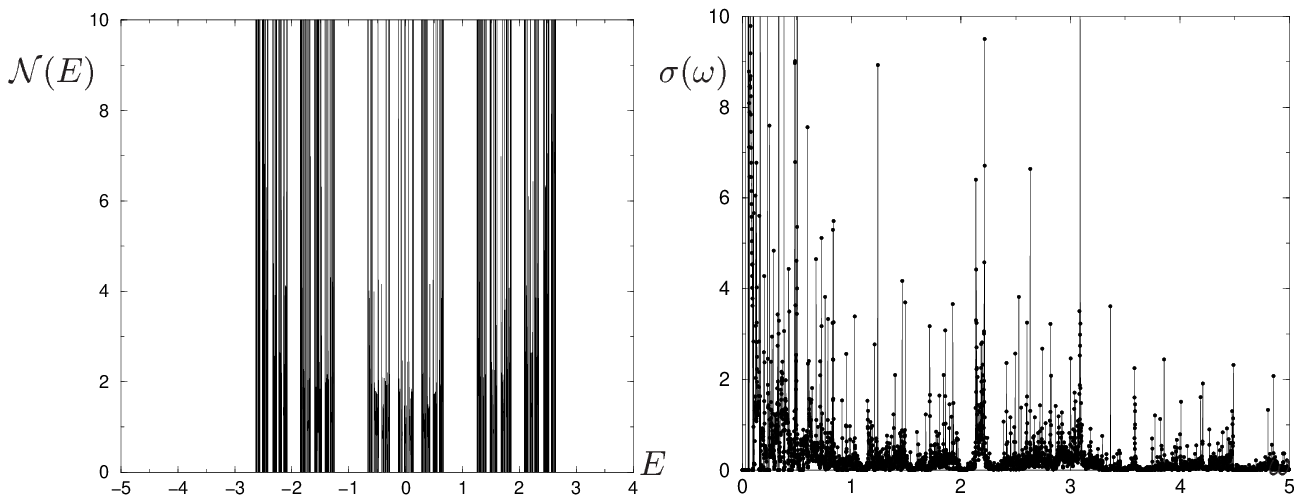}}

  \caption{Density of states ${\cal N}(E)$ (left) and conductivity 
           $\sigma(\omega)$ (right) of the 
           FC with a supercell of $f_{11} = 233$ sites. 
           Fermi energy $\epsilon_{F} = 0$,
           $t_S = 1.0, t_L = 1.5$, and
           $\epsilon_{\alpha} = \epsilon_{\beta} = \epsilon_{\gamma} = 0$  
           (arbitrary units).}
  \label{sig1-fig} 

  \vspace*{24pt} 
  \centerline{\includegraphics[width=5in]{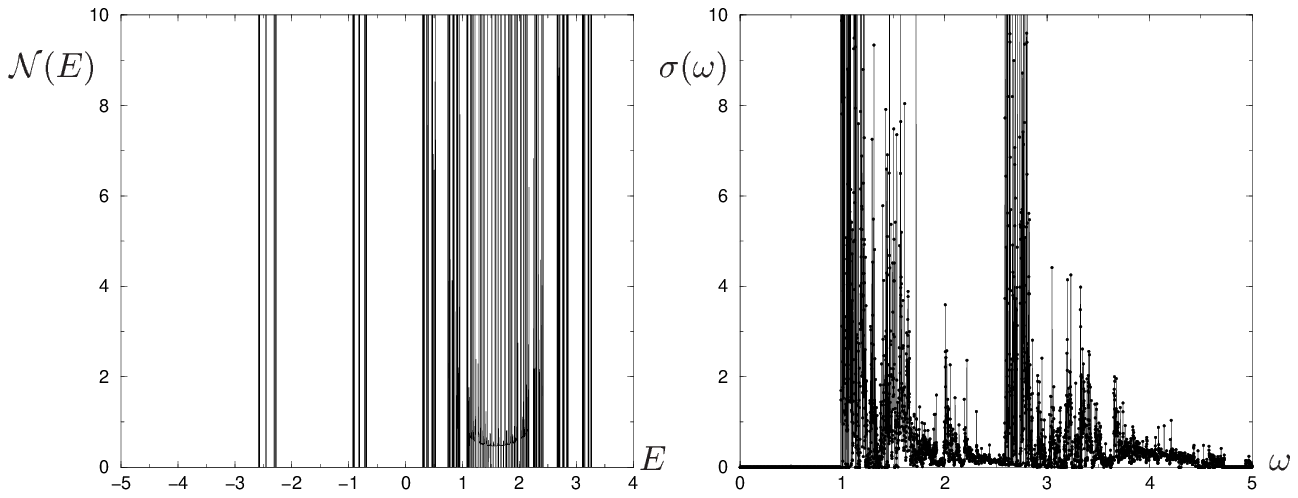}}

  \caption{Density of states (left) and conductivity (right) of the  
           same FC as in Fig.\ \ref{sig1-fig} but with different parameters:
            $\epsilon_{F} = -0.2$,
            $- \epsilon_{\alpha} = \epsilon_{\beta} = \epsilon_{\gamma} = 1.0$.}
  \label{sig2-fig}
\end{figure}

Figs.\ \ref{DOS-fig}, \ref{sig1-fig}, \ref{sig2-fig} display some numerical 
results for the DOS, mini--{}\!\! bandstructure, transition energies,  and the conductivity.
We note, that the expression of $\sigma^{(n)}(\omega)$ in Eq.\ (\ref{pr20}) is nonzero only for $\omega > 0$.
For  $\omega \rightarrow 0$,  merely $\sigma^{(n)}(0)$ contributes because none of the
lines  $(E^{(i)}_{\omega}, \omega)$ cut the axis $\omega=0$ for energies {\it within} the spectrum.
Therefore, $\sigma^{(n)}(\omega)$ vanishes in the limit $\omega \rightarrow 0$.
(In addition, the $dc$ conductivity (see after Eq.~(\ref{pr20})) 
vanishes in the limit of infinite aperiodic FC, $n\to\infty$.)
This is characteristic for the (infinite) TB model, as $\HO_{_{\Sigma_n}}$ is a 
bounded operator. Thus, there is always a gap in the neighbourhood of $\omega=0$. 
If one introduces some mechanism of dissipation or disorder one may expect that
the small frequency--gap revealed in Fig.\ \ref{sig1-fig} may be smeared out 
(leading to metallic--like behaviour),  but will persist in the case of
Fig.\ \ref{sig2-fig} (semiconductor--like behaviour). 

The main numerical challenge turned out to be the calculation of the
energies $E_{\omega}$, i.e., the zeros of Eq.~(\ref{tr-en}).
This task implicitly implies polynomials of very high order. Such operations
are numerically unstable but could be managed with the help of  Bailey's
multiprecision software packet\cite{Bailey}.

Starting from the self--similarity of the FC we found a real space 
renormalization scheme for the Kubo--conductivity which is suitable 
for numerical implementation. 
Hereto, this scheme stands alongside with the powerful tool of the 
trace map for the spectrum\cite{Kohmoto} and
the path renormalisation scheme for the local DOS\cite{Walther}.
For details and generalizations of our approach to arbitrary morphisms 
(like Thue--Morse, Rudin--Shapiro, period doubling) and
n-letter alphabets, we refer to Ref.\ \onlinecite{WaltherDiss}.

\section*{ACKNOWLEDGMENTS}

This work was supported by the Deutsche Forschungsgemeinschaft DFG
through the ``Schwerpunktsprogramm''  {\it Quasikristalle}. 
D. H. Bailey and NASA Ames are kindly acknowledged for making 
available their {\it Multiple Precision Floating Point 
Computation Package}\cite{Bailey}.

\end{document}